\documentclass[epj, referee]{svjour}
\usepackage{graphics}
\begin{document}
\title{Longitudinal Scaling of Elliptic Flow in Landau Hydrodynamics}

\author{Karolis Tamosiunas\inst{1} \inst{2} 
\thanks{karolis.tamosiunas@tfai.vu.lt}%
}                     
\offprints{}          
\institute{ Vanderbilt University, Nashville, Tennessee 37235, USA \and Institute of Theoretical Physics and Astronomy, Vilnius University, Go\v stauto g. 12, 01108 Vilnius, Lithuania}
\date{Received: date / Revised version: date}
%
\abstract{
This study presents a generalization of the Landau hydrodynamic solution for multiparticle production applied to non-central relativistic heavy ion collisions. The obtained results show longitudinal scaling of elliptic flow, $v_2$, as a function of rapidity shifted by beam rapidity ($y-y_{beam}$) for different energies ($\sqrt{s_{NN}}=62.4$ GeV and $200$ GeV) and for different systems (Au-Au and Cu-Cu). It is argued that the elliptic flow and its longitudinal scaling are due to the initial transverse energy density distribution and initial longitudinal thickness effect. 
} 
\newcommand{\beq}{\begin{equation}}
\newcommand{\eeq}[1]{\label{#1} \end{equation}}
\newcommand{\bdm}{\begin{displaymath}}
\newcommand{\edm}{\end{displaymath}}
\maketitle
\section{Introduction}

 \label{intro}
Experimentally observed azimuthal asymmetries of particle production in non-central heavy ion collisions are currently of high interest, as they provide more information about the early dynamics of the high-energy nuclear reactions. Moreover, RHIC data \cite{rhic-data05} of elliptic flow for different pseudorapidities shows universal behavior for different nuclei and for different beam energies. Not all of the theoretical models are able to reproduce the observed longitudinal scaling of elliptic flow \cite{long-scale}. For example, the AMPT (a multiphase transport model, ver 1.11) \cite{AMPT} and UrQMD (ultrarelativistic quantum molecular dynamics, ver 2.3) \cite{UrQMD} do not reproduce experimentally observed scaling of elliptic flow. While, the AMPT with string melting and the Buda-Lund model, based on an analytic solution of perfect fluid dynamics, can reproduce experimental data \cite{v2scaleBL}.

Landau's approximate hydrodynamic solution for particle production in relativistic heavy ion reactions \cite{Landau56,Landau53} was formulated in 1953, but even today gets a lot of attention and reproduces well the rapidity spectrum of observed particles in the relativistic nuclear collisions  \cite{Steinberg:2004vy}. In recent studies  \cite{Csernai10,Wong:2008ex} the particle production function for different rapidities was slightly modified compared to the original version to the following form: 
\beq
dN/dy \propto \exp \sqrt{(y_{beam}^2-y^2)},  
\eeq{dndy}
where $y_{beam}=\ln(\sqrt{s_{NN}}/m_N)$ is the beam rapidity. 
In order to make a comparison with experimental particle spectra one has to normalize the distribution with the total number of particles, which is unknown from the Landau solution. The key reason lies in the fact that the Landau solution does not conserve total energy, thus the total entropy of the system cannot be obtained which leads to an unknown number of produced particles $N_{tot}$. Anyway, the solution of perfect fluid dynamics (\ref{dndy}) gives the right shape of the rapidity distribution for different relativistic energies. 
On the other hand, from the definition of the elliptic flow:   
\beq
v_2(y)=\frac{\int d\phi (dN/d\phi dy) \cos(2\phi)}{\int d\phi (dN/d\phi dy) }, 
\eeq{v2}
one can easily see that normalization of total particle production is not needed. So the task is to modify the approximate solution of Landau, including the transverse angle $\phi$ dependence in the solution.    

In this study the analytic solution for relativistic hydrodynamic equations will be presented. In section \ref{sec-long} the longitudinal expansion will be summarized for the sake of completeness of the study, even though the same results can be found elsewhere. The transverse part of the solution with asymmetrical pressure gradient driven expansion are presented in section \ref{sec-trans}. The last section presents initial state and obtained results. 
 
The presented solution is approximate, but in comparison to the computational hydrodynamics is analytic and transparent. The main assumptions of the model coincides with original Landau approximations and are as follows: 
i) longitudinal and transverse parts of hydrodynamic equations are solved separately; 
ii) the equation of state of ideal relativistic gas, $P=e/3$, is used to solve transport equations;
iii) transverse expansion does not include initial flow and is pressure gradient driven. 

\section{Longitudinal Expansion}\label{sec-long}

In this model we solve the equations of energy-momentum conservation:
\beq
\partial_\mu T^{\mu\nu}=0 ,
\eeq{hydroT}
where the energy-momentum tensor reads as: 
\beq
T^{\mu\nu}=(e+P)u^\mu u^\nu -Pg^{\mu \nu}.
\eeq{def_T4}
The solution of local conservation laws (eq. \ref{hydroT}) with the equation of state represents the dynamics of the system by relating bulk properties of the matter, such as: energy density, $e$, local pressure, $P$ and the four-flow of the fluid, $u^\mu=u^0(1, \vec v)$. 
The equations of the hydrodynamic longitudinal expansion in 1+1 dimension, along the $z$ axis reads as: 
\beq 
\frac {\partial T^{00}}{\partial t} + \frac{\partial T^{0z}}{\partial z}=0 , \  \
\frac {\partial T^{0z}}{\partial t} + \frac{\partial T^{zz}}{\partial z}=0.
\eeq{long2}
The details on how to solve the above equations can be found in \cite{Landau56}, \cite{Landau53},   \cite{Csernai10}, \cite{Wong:2008ex}, \cite{Khalatnikov54}, \cite{Khalatnikov04}, so here we present the derivation shortly including only main equations and the result, as we will need it later.  

Solution of the equations of hydrodynamics (\ref{long2}) starts by transforming relativistic velocity field components to rapidity terms, as:
\beq
u^0=\cosh y , \  \   u^z=\sinh y .
\eeq{u_to_rap}
From the above transformation, naturally follows, that: $v_z=\tanh y$ and $(u^z)^2 - (u^0)^2=-1$. 
In order to solve hydrodynamic equations (\ref{long2}), the following variables are introduced: 
$$ \kappa=\ln w , \ \ w=e+p ,\ \ \chi = \psi - wu_zz - wu_0t , $$ where $\chi$ is called the Khalatvikov potential, $\psi$ is hydrodynamic potential, which is defined by the relation: $
wu_i=\frac{\partial \psi}{\partial x^i} $, where $\psi$ is a function of coordinates and time. And, $w=e+P$ is the enthalpy. 
Using the relation for sound velocity, as:
 $$ c_s^2=\frac{n}{w}\frac{\partial w}{\partial n} ,$$
 where $c^2_s=1/3$ for the ideal relativistic gas equation of state \cite{LLfluid}. After the Legendre transformation to the hodograph plane, the Chaplygin equation for supersonic expansion reads as: 
 \beq
 c_s^2\frac{\partial ^2 \chi}{\partial \kappa ^2} +(1-c_s^2)\frac{\partial  \chi}{\partial \kappa } -\frac{\partial ^2 \chi}{\partial y ^2}=0 .
 \eeq{Chap} 
Detailed investigation on the solution of the 1+1 dimensional perfect fluid hydrodynamics can be found in \cite{beuf08,sarid11,Mizo}.  The solution for above equation reads:
\begin{eqnarray}
t=e^{-\kappa} \Big(\frac{\partial \chi}{\partial \kappa} \cosh y - \frac{\partial \chi}{\partial y} \sinh y  \Big) , \\
z=e^{-\kappa} \Big(\frac{\partial \chi}{\partial \kappa} \sinh y - \frac{\partial \chi}{\partial y} \cosh y  \Big). 
\end{eqnarray}\label{pot_sol}
The transformation back to the $(t ,z)$ coordinates is shorter with the following new variables: 
\beq
y_+= \ln((t+z)/\Delta), \  \
y_-=\ln((t-z)/\Delta),  
\eeq{y+-}
where $\Delta$ is the initial thickness of the system in the beam direction, $z$. Also, $\Delta$ is the initial condition after which the equation of state is assumed to be valid and evolution equations (\ref{long2}) are applied.

The final solution for energy density, $e(y_+,y_-)$, and rapidity, $y(y_+,y_-)$, is: 
\beq
e(y_+,y_-)=e_0 \exp[-4/3(y_+ + y_- -\sqrt {y_+ y_-})] ,
\eeq{en_dnz}
\beq
y(y_+,y_-)= ( y_+ - y_-)/2 , 
\eeq{z_y}
where $z=t\tanh y$. 
The above solution of  1+1-dimensional relativistic hydrodynamics equation (\ref{long2}) will be connected to the solution of transverse expansion, in order to obtain multiplicities of produced particles for different rapidities. 

\section{Transverse expansion} \label{sec-trans}

In order to solve the transverse part of hydrodynamic equations (\ref{hydroT}) we will follow original Landau assumptions with some modifications. For simplicity, polar coordinates will be used, where four-flow components in polar coordinates are: $
  u_i={dx_i}/{dt}$, $ u_0=(1-(\frac{d}{dt} r^2 + r^2 \frac{d}{dt} \phi^2))^{-1/2}$, $u_r=u_0v_r$ and energy-momentum tensor (\ref{def_T4}) components are as follows:  
 $$
T^{rr}= (e+P) (u^0)^2v_r^2+P, \ \  T^{\phi\phi}=(e+P)(u^0)^2v_\phi^2 + P/r^2
$$
$$
T^{0r}=(e+P) (u^0)^2v_r, \ \ T^{0\phi}=(e+P) (u^0)^2v_\phi \ .
$$
Assuming that the transverse velocity is radial, $v_\phi=0$, the hydrodynamic equation (\ref{hydroT}) for the transverse dynamics at  the fixed transverse angle $\phi$ becomes: 
  \beq
\frac {\partial T^{0r}}{\partial t}  + \frac {\partial T^{rr} }{\partial r}=0 \ .
\eeq{tr_p}
Inserting energy-momentum tensor expressions in the above equation and using ideal gas equation of state, $P=e/3$, one gets: 
\beq
4e(u^0)^2\frac {\partial v_r}{\partial t} +4e(u^0)^2 \frac {\partial v_r^2 }{\partial r} +\frac {\partial e }{\partial r}=0 \ .
\eeq{tra1}
Following the original Landau derivation, the fist term in the above equation is an acceleration dependence, and is assumed to be equal to ${\partial v_r}/{\partial t} =2 r(t)/t^2$. The second term is set to zero,  $v_r$ being comparatively small. To simplify the third term, Landau used $\partial e /\partial r \approx - e/R_A$, because the energy density at the center has value $e$ and is zero at the edge of the system, $r=R_A$. In the case of peripheral collisions, we do not expect centrally symmetric energy density distribution, thus the assumption is modified, as: 
\beq
\frac{\partial e}{\partial r}=\frac{e(r=R_\phi)-e(r=0)}{R_\phi} ,
\eeq{tra2}
where $R_\phi$ is a transverse radius of the system, which changes with the angle, as the system is not centrally symmetric. We do not know the value of $e(R_\phi)$, so we introduce a new function, $f(R_\phi)=e(r=R_\phi)/e(r=0)$, which is a fraction of energy density at the edge of the system with respect to the energy density at the center.  
In this way, the function $f(R_\phi)$ must be less than unity for any angle $\phi$, as the energy density at the center is higher than at the edge. This modification from the original Landau assumption plays an important role, as it involves transverse asymmetry to the solution by making a different transverse pressure gradient for different  $\phi$ angles.
  Now from eq. (\ref{tra1}), we express the transverse displacement dependence on time, as:  
\beq
r(t)=\frac{(1-f(R_\phi))t^2}{8(u^0)^2 R_\phi} .
\eeq{rad}
The last stage of the model is the so called conic-flight stage, where energy and entropy fluxes stop changing for the fixed cone element $2 \pi r dr$. This stage coincides with the kinetic Freeze-Out (FO), as the model does not include any hadronic re-interactions after the evolution stops.    With the help of the formula (\ref{rad}) we obtain a hypersurface in space-time, after which hydrodynamic flow stops and matter streams freely towards detectors. Here again we follow the original Landau model, assuming a fixed transverse distance, $r(t_{FO})=a=2R_A$, for the conic-flight to start at the distance of the nuclear  diameter. Thus, we obtain the value for the time, when the conic-flight starts, which reads as:  
\beq
t_{FO}=2\cosh y \sqrt{\frac{2aR_\phi}{(1-f(R_\phi))}},  
\eeq{time}
where the relation $u^0=\cosh y$ was used.

The solution in the conic-flight stage is straightforward, as the energy and entropy does not change at a fixed cone element. The transverse and longitudinal solutions are matched at the time $t=t_{FO}$. Knowing that  $dS=s u^0 dz$ at a given time within element $dz$, and that the entropy density is $s=ce^{3/4}$, we can express the entropy change over rapidity from the energy density formula (\ref{en_dnz}) as:  
\beq
\frac{dS}{dy}=ce_0^{3/4} \exp[-(y_+  + y_- -\sqrt {y_+ y_-})]\frac{t}{\cosh y}. 
\eeq{entr}
Inserting the solution for the FO time equation (\ref{time}) into the entropy equation above and assuming that the number of produced particles is directly proportional to the entropy,  $dN \propto dS$, one can obtain the number of particles for different rapidities at a fixed angle $\phi$. However,   the function $f(R_\phi)$ and initial thickness $\Delta$ is still needed. 
 
 \section{Initial conditions and results} \label{sec-init}
 
 As is natural for hydrodynamics, the initial state is based on the predictions from other models. In this case the widely accepted and analytically simple Wounded Nucleon (WN) model \cite{wn} will be used to parametrize initial conditions. It is based on the Woods-Saxon nuclear density parametrization \cite{woodssaxon}, as follows: 
\beq
\rho_A({\bf r})=\frac{\rho_0}{1+\exp(\frac{{\bf r}-R_A}{d})} ,
\eeq{WS}
 which is continuous and can be straightforwardly connected to the Landau equations. The main requirement for the initial conditions and new function $f(R_\phi)$ is that for the central collision case, $b=0$, the result must be equal to the original Landau one. Moreover, the WN model connects the impact parameter $b$ with the number of participating nucleons, $N_{part}$, and the number of binary collisoins, $N_{coll}$, making comparisons with the experimental data easy.  
 
The density of wounded nucleons in the transverse plane and in polar coordinates, $(r,\phi)$, can be obtained by: 
\begin{eqnarray*} 
n_{WN}(r,\phi)&=&T_A(r,\phi)\Bigg[ 1-\Big(1-\frac{\sigma T_B(r,\phi)}{B} \Big)^B\Bigg] \\ &+& T_B(r,\phi)\Bigg[ 1-\Big(1-\frac{\sigma T_A(r,\phi)}{A} \Big)^A\Bigg] .
\end{eqnarray*}\label{dens_wn}
Here and everywhere else the vector $r$ starts at the center of the almond shaped system of interest. 
The thickness functions are then expressed as: 
$$
T_A(r,\phi)=T_A(x-b/2,y)=\int  dz \  \rho_A(\bf r),
$$
$$
T_B(r,\phi)=T_B(x+b/2,y)=\int  dz \  \rho_B(\bf r),
$$
using a Woods-Saxon parametrization (\ref{WS}) with $R_A=1.12A^{1/3}-0.86A^{-1/3}$ [fm], $d=0.54$ [fm] and $n_0=0.17 fm^{-3}$.  

Now assuming that the energy density is proportional to the WN density \cite{thick}: $e(r, \phi; b) \propto n_{WN}(r,\phi ;b)$, the function $f(R_\phi)$ can be obtained. It is, by definition, the ratio of energy density at the edge of the system to the energy density at the center, for a fixed impact parameter $b$ and reads as: 
\beq
f(R_\phi)=\frac{n_{WN}(R_\phi, \phi ;b)-\min(n_{WN}(R_{\phi}, \phi ;b))}{n_{WN}(0,0;b)} .  
\eeq{fnuor}
The radius of the system, $R_\phi$, is dependent on the angle $\phi$ and is obtained from the geometry of two overlapping circles as: 
$R_\phi^2 + R_\phi b\cos{\phi}+\frac{b^2}{4}-R_A^2=0$. 
The term $\min(n_{WN}(R_{\phi}, \phi ;b))$ is a minimal density at the edge of the system and is used in order to have the original Landau solution for $R_\phi=R_A$, so that $f(R_A)=0$. Because  Landau originally used a "sharp sphere" picture, assuming that the energy density is zero at the edge of the system while solving (\ref{tra1}). In the case of the Woods-Saxon model, the density (\ref{WS}) at the edge of the nucleus at $r=R_A$ is not zero, so the minimal value is subtracted. Now the acceleration term in (\ref{tra1}) is the same in central collision and in Landau, but for peripheral collisions the acceleration does depend on the angle $\phi$. 

\begin{figure}
\resizebox{1.0\columnwidth}{!}{%
  \includegraphics{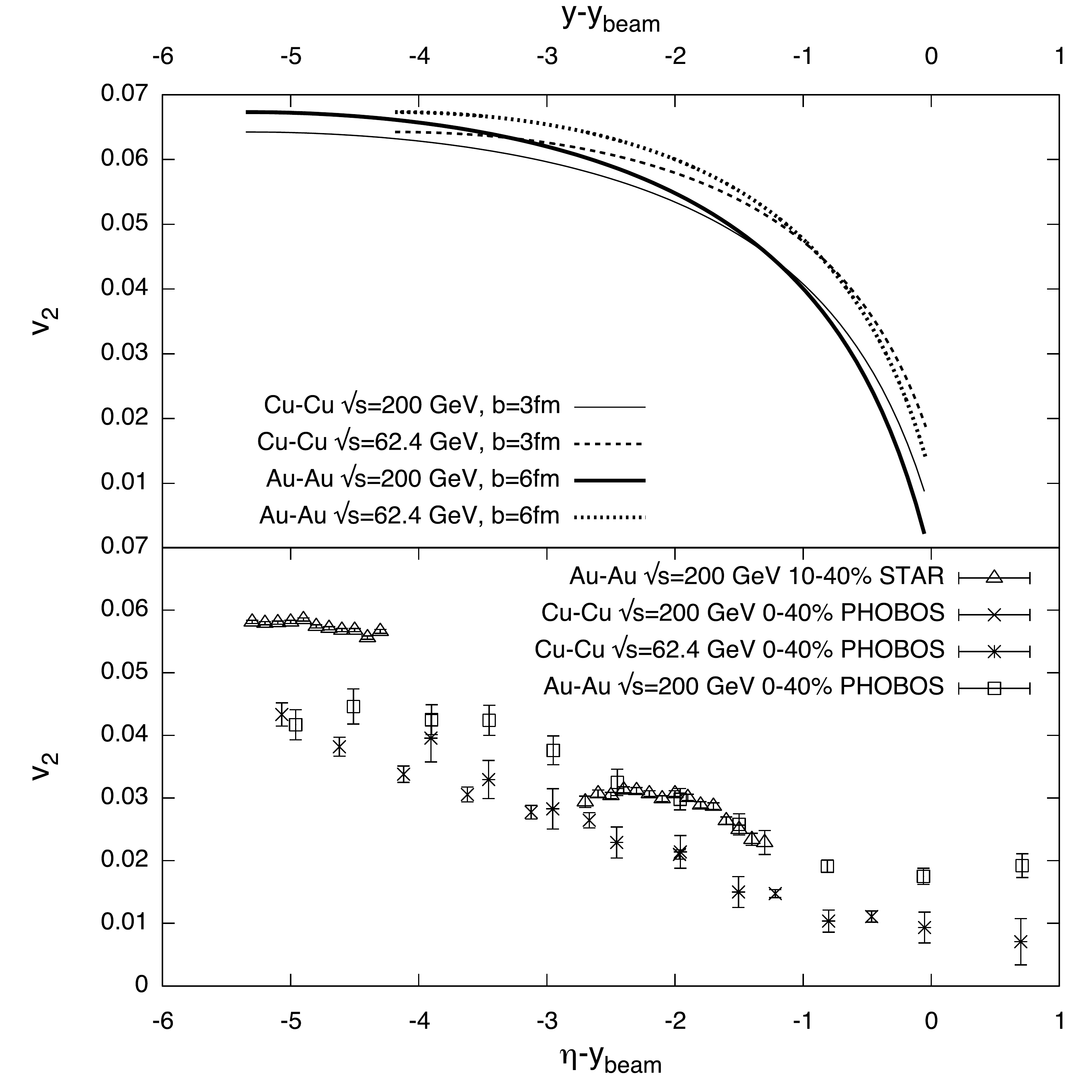}
}
\caption{Elliptic flow dependance on (pseudo)rapidity for Au-Au and Cu-Cu for two different collision energies $\sqrt{s_{NN}}=62.4$ GeV and $200$ GeV and for different centralities as indicated on the figure.The model outcome is on the top, while data from RHIC experiments \cite{star-data} and \cite{rap_dist_rhic} are on the bottom. }
 \label{fig1}       
\end{figure}

Finally, to calculate the elliptic flow  (\ref{v2}) one should merge equations (\ref{y+-}, \ref{time}, \ref{entr}, \ref{fnuor}) and the initial longitudinal thickness $\Delta$, which for the peripheral collisions is expressed as: 
\beq
\Delta(\phi)=\kappa_\phi R_A/\gamma ,
\eeq{delta}
where $\kappa_\phi=\sqrt{n_{WN}(R_\phi, \phi ;b)/\max(n_{WN}(R_{\phi}, \phi ;b))}$. It means that the longitudinal expansion (\ref{long2}) starts with an azimuthally asymmetric initial thickness, which is wider where the initial nuclear density is higher. 
The term $\max(n_{WN}(R_{\phi}, \phi ;b))$ is used in order to have $\kappa_\phi(b=0)=1$ for the central collision case. The energy dependance on the initial state is found via the Lorentz gamma factor, $\gamma=\sqrt{s_{NN}}/2m_{proton}$. The obtained results are shown in figure (\ref{fig1}) for reactions of Au-Au at $b=6$fm and Cu-Cu at $b=3$fm for two different beam energies, as observed at RHIC. One can easily see that the model does predict the longitudinal scaling, but does not correspond well to the experimental data. The latter can be explained by the lack of realistic Freeze-Out dynamics and hadronic reinteractions in the model. Moreover, the initial state is chosen to be as simple and transparent as possible, in order to show pure hydrodynamic effects for a system with transverse asymmetry and does not include initial state fluctuations. 
The initial energy density distribution $f(R_\phi)$ and initial longitudinal thickness $\Delta(\phi)$ might be based on more sophisticated models, but for now we can conclude that Landau hydrodynamic solution and it's assumptions work not only for particle multiplicity spectra, but for elliptic flow and elliptic flow longitudinal scaling as well.

The generalized solution of Landau hydrodynamics incorporates transverse asymmetries into the initial configuration and can be compared to a wide amount of experimental data, giving deeper insight into the early dynamics of the system. 

{\bf Acknowledgement} 
K.T. acknowledges the support from the Baltic-American Freedom Foundation and Vanderbilt University Relativistic Heavy Ion Group for valuable comments.


\begin{thebibliography}{}
\bibitem{rhic-data05} B.B. Back, et al, Phys. Rev. Lett. {\bf 94} (2005), 122303.
\bibitem{long-scale} Md. Nasim et al, Phys. Rev. C {\bf 83} (2011) 054902.
\bibitem{AMPT} Zi-wei Lin and C. M. Ko, Phys. Rev. C {bf 65}, (2002) 034904.
\bibitem{UrQMD} M. Bleicher, et al, J. Phys. G: Nucl. Part. Phys. {\bf 25} (1999) 1859Ð1896.
\bibitem{v2scaleBL} M. Csanad, et al, Eur. Phys. J. A {\bf 38} (2008) 363-368.
\bibitem{Landau56}
  S.~Z.~Belenkij and L.~D.~Landau,
  Nuovo Cim.\ Suppl.\  {\bf 3S10} (1956) 15
  [Usp.\ Fiz.\ Nauk {\bf 56} (1955) 309].
\bibitem {Landau53} L.D. Landau Izv. Akad. Nauk SSSR {\bf 17} 51 (1953)  [English translation: \textit{Collected Papers of L. D. Landau}, edited by D. ter Haar, Gordon and Breach, New-York, (1968)].
\bibitem{Steinberg:2004vy}  P. Steinberg,
  Acta Phys.\ Hung.\  A {\bf 24} (2005) 51
  [arXiv:nucl-ex/0405022].
   \bibitem{Csernai10} M. Zetenyi, L.P. Csernai, Phys.\ Rev.\ C {\bf 81} (2010) 044908.
\bibitem{Wong:2008ex}  C. Y. Wong,  Phys. Rev.  C {\bf 78} (2008) 054902, [arXiv:0808.1294 [hep-ph]].
  \bibitem{star-data} B. I. Abelev, et al, Phys. Rev. C 77 (2008) 54901.
  \bibitem{rap_dist_rhic} B. Alver, et al, for the PHOBOS Collaboration, Phys. Rev. Lett. {\bf 98} (2007) 242302.
\bibitem{Khalatnikov54} I.M. Khalatnikov, Zh. Eksp. Teor. Fiz. 27 (1954) 529.
\bibitem{Khalatnikov04} I.M. Khalatnikov, A.Yu. Kamneshchik, Phys. Lett. A {\bf 331} (2004) 12-14.
  \bibitem{LLfluid} L.D. Landau, E.M. Lifshitz, \textit{Fluid Mechanics}, Pergamon press, Oxford, UK, 1987.
 \bibitem{beuf08} G. Beuf, R. Peschanski, and E. Saridakis, Phys. Rev. C {\bf 78}, (2008) 064909  [arXiv:0808.1073].
 \bibitem{sarid11} R. Peschanski, E.N. Saridakis, Nucl. Phys. A {\bf 849} (2011) 147-164.
 \bibitem{Mizo} T. Mizoguchi, H. Miyazawa and M. Biyajima, Eur. Phys. J. A {\bf 40} (2009) 99-108 
\bibitem{wn} A. Bialas, M. Bleszynski, W. Czyz, Nucl. Phys. B {\bf 111} (1976) 461.
\bibitem{woodssaxon} R.D. Woods, D.S. Saxon, Phys. Rev. {\bf 95}, (1954) 577, doi:10.1103/PhysRev.95.577.
\bibitem{thick} P.F. Kolb et al, Nuclear Physics A  {\bf 696} (2001) 197Ð215.




\end{thebibliography}
\end{document}